 \documentclass[manuscript,screen]{acmart}

\AtBeginDocument{%
  \providecommand\BibTeX{{%
    \normalfont B\kern-0.5em{\scshape i\kern-0.25em b}\kern-0.8em\TeX}}}

\setcopyright{acmcopyright}
\copyrightyear{2021}
\acmYear{2021}
\acmDOI{10.1145/1122445.1122456}

\acmConference[CHI '21]{CHI '21: ACM Conference on Human Factors in Computing Systems}{May 08--13, 2021}{Yokohama, Japan}
\acmBooktitle{CHI '21: ACM Conference on Human Factors in Computing Systems,
  May 08--13, 2021, Yokohama, Japan}
\acmPrice{15.00}
\acmISBN{978-1-4503-XXXX-X/18/06}

\begin{document}

\title{Look Before You Leap: Trusted User Interfaces for the Immersive Web}


\author{Diane Hosfelt}
\affiliation{\institution{Mozilla}}
\email{dhosfelt@mozilla.com}
\author{Jessica Outlaw}
\affiliation{\institution{The Extended Mind}}
\author{Tyesha Snow}
\affiliation{\institution{The Extended Mind}}
\author{Sara Carbonneau}
\affiliation{\institution{The Extended Mind}}

\renewcommand{\shortauthors}{Hosfelt}

\begin{abstract} % max 150 words
Part of what makes the web successful is that anyone can publish content and browsers maintain certain safety guarantees. For example, it’s safe to travel between links and make other trust decisions on the web because users can always identify the location they are at. If we want virtual and augmented reality to be successful, we need that same safety. On the traditional, two-dimensional (2D) web, this user interface (UI) is provided by the browser bars and borders (also known as the chrome). However, the immersive, three-dimensional (3D) web has no concept of a browser chrome, preventing routine user inspection of URLs. In this paper, we discuss the unique challenges that fully immersive head-worn computing devices provide to this model, evaluate three different strategies for trusted immersive UI, and make specific recommendations to increase user safety and reduce the risks of spoofing.
\end{abstract}

\begin{CCSXML}
<ccs2012>
<concept>
<concept_id>10002978.10003029.10011703</concept_id>
<concept_desc>Security and privacy~Usability in security and privacy</concept_desc>
<concept_significance>500</concept_significance>
</concept>
<concept>
<concept_id>10002978.10003006.10003011</concept_id>
<concept_desc>Security and privacy~Browser security</concept_desc>
<concept_significance>300</concept_significance>
</concept>
<concept>
 <concept>
<concept_id>10003120.10003121.10003124.10010392</concept_id>
<concept_desc>Human-centered computing~Mixed / augmented reality</concept_desc>
<concept_significance>100</concept_significance>
</concept>
</ccs2012>
\end{CCSXML}

\ccsdesc[500]{Security and privacy~Usability in security and privacy}
\ccsdesc[300]{Security and privacy~Browser security}
\ccsdesc[100]{Human-centered computing~Mixed / augmented reality}
\keywords{link traversal, trusted user interfaces, immersive web}


\begin{teaserfigure}
  \includegraphics[width=\textwidth]{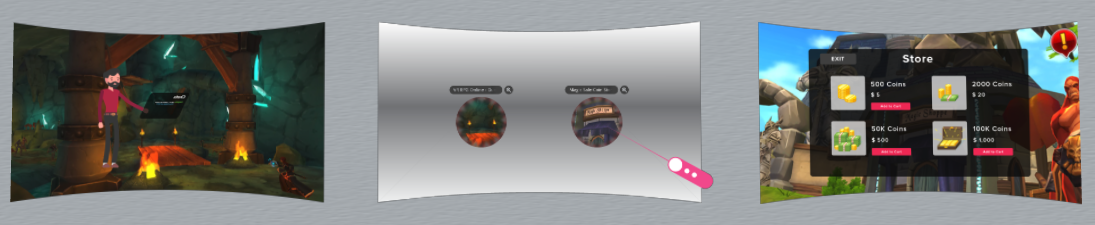}
  \caption{Basic immersive link navigation flow without trusted UI}
  \Description{Immersive link navigation flow without trusted UI}
  \label{fig:linktraversal}
\end{teaserfigure}

\maketitle

\section{Introduction}
Browsing the web is a common, everyday task. Most people traverse links without thinking about the safety mechanisms that enable them to travel from one web location to another. Key among these is the fact that at any point in time, you can inspect your location by looking at the URL bar. In order to make trust decisions, users must first understand the identity of the website they are visiting, which is indicated by the URL and HTTPS certificate data displayed in the URL bar. However, previous research indicates that users often fail to note these indicators~\cite{dhamija2006phishing,thompson2019web,lin2011does}.

\begin{figure}[ht]
    \centering
    \begin{minipage}[b]{0.45\linewidth}
    \includegraphics[width=\linewidth]{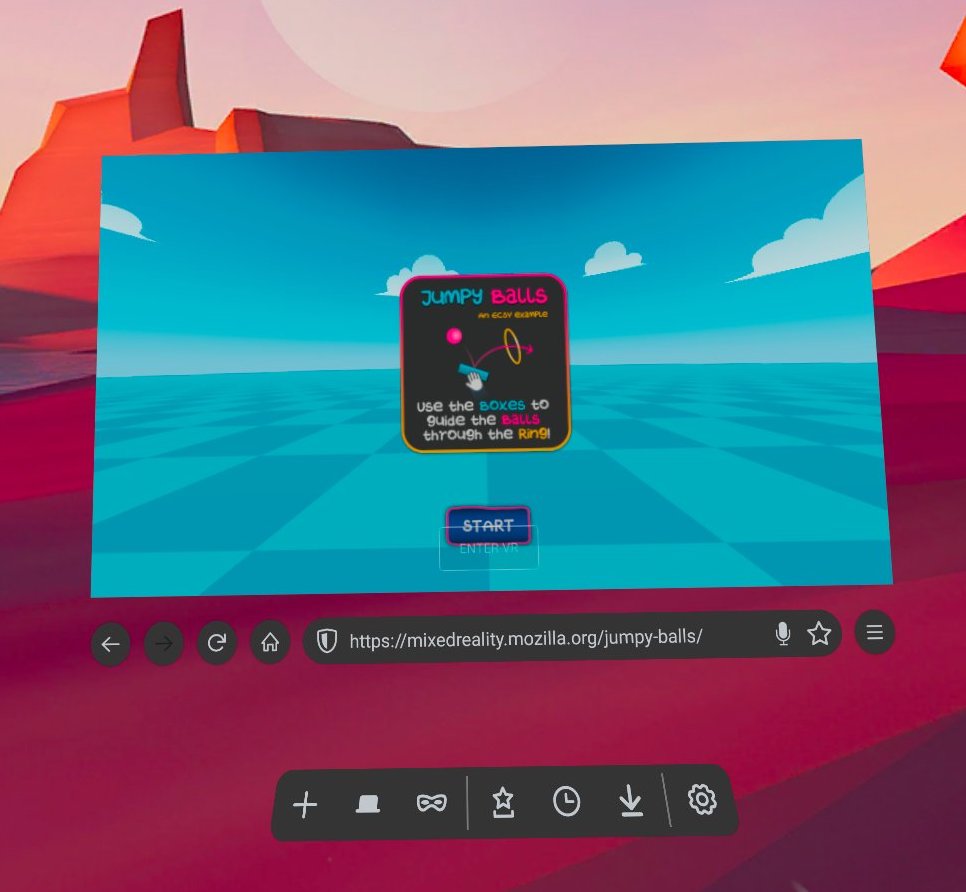}
    \caption{A screenshot of 2D mode on the Oculus Quest, where the URL bar and other browser chrome elements are visible. The 'Enter VR' button is shown just below the 'Start' button.}
    \label{fig:2d}
    \end{minipage}
    \quad
    \begin{minipage}[b]{0.45\linewidth}
    \includegraphics[width=\linewidth]{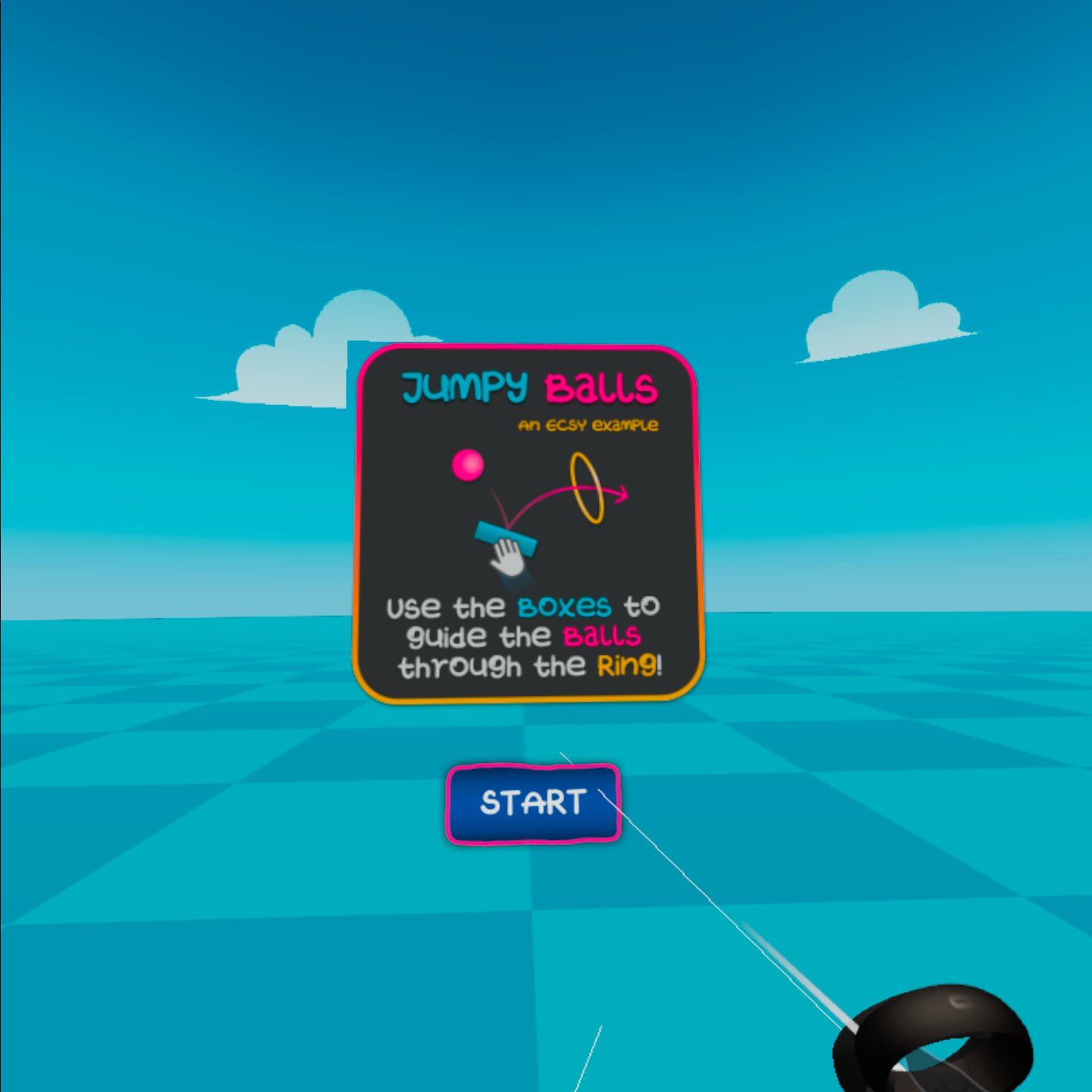}
    \caption{A screenshot of immersive mode on the Oculus Quest, where the webpage renders all of the pixels. This mode is accessed by pressing the 'Enter VR' button shown in \autoref{fig:2d}}
    \label{fig:immersive}
    \end{minipage}
\end{figure}

The \emph{immersive web} refers to virtual experiences hosted through the browser~\cite{immersive-web}. This experience is commonly accessed on a head-mounted virtual reality (VR) device like an Oculus Quest or HTC Vive. Immersive web browsers are capable of displaying 3D content to users, as well as immersive 360\textdegree~videos, and augmented reality (AR) content on AR-enabled devices. The immersive web has two modes, 2D and immersive (shown in \autoref{fig:2d} and \autoref{fig:immersive} respectively), where the 2D mode is simply a 2D browser embedded in a 3D environment and immersive mode is a fully immersive experience. In this context, \emph{immersive} means to fully engage the user's visual sense and present an experience that appears to surround the user. Following Milgram's taxonomy, this paper is focused on immersive virtual reality experiences~\cite{milgram1994taxonomy}.

On a 2D browser, the browser chrome is the borders of the browser window, including the window frames, menus, toolbars, and scroll bars. Only the browser can render the browser chrome, including the URL bar. In immersive mode, the immersive web lacks a browser chrome. The browser chrome prevents spoofing, because new windows and tabs always have the full browser chrome---that is, developers can't interfere with anything above the web page window. If a browser UI can be influenced by arbitrary web sites to hide the URL or modify how it is displayed, then a malicious web site can spoof UI elements to display arbitrary URLs, tricking the user into thinking they are browsing a trusted site.

When a user wishes to know where they are in a 2D browser, it is as simple as looking at the URL bar, but there is no obvious way to do this in 3D immersive mode. Similarly, traversing a link safely requires that users be able to understand and inspect the location of where they are going, which is possible in 2D, and not in 3D. This is because browsing on the immersive web is most similar to the 2D analogue of full-screen mode---the web page controls all of the pixels that a user sees. On the 2D web, it is reasonable to ask users to exit full-screen mode to navigate to a new site. However, leaving immersive mode is particularly jarring in a 3D environment (and could be disorienting if a user must remove a headset to do so), serving as a barrier to adoption, as well as potentially breaking sites.

To follow a link in immersive mode on the immersive web, currently a user must exit immersive mode to 2D mode, follow the link in 2D mode, then re-enter immersive mode on the new web page. A key goal for the immersive web is to enable immersive-to-immersive navigation that does not require an exit to 2D mode. To do so there are a number of problems that must be solved first, namely the lack of a link standard for the immersive web, and the inability to inspect URLs to protect against redirection attacks and phishing. This paper addresses the latter.

Drawing inspiration from the browser chrome, the solution to this problem is to create a mechanism to indicate that the browser is rendering the content being shown. This method enables the following activities:
\begin{itemize}
    \item Link traversal
    \item Inspecting URLs
    \item Accountability (inspecting and revoking consent)\footnote{Granting consent does not require this mechanism.}
\end{itemize}

In this paper, we propose three different security user interface (UI) mechanisms for indicating that a browser is rendering the content being shown---in this case, the content being the destination of a link traversal. These mechanisms allow users to make an informed decision if they would like to follow the link or return to their current location. We then examine the effectiveness of each proposal via a user study, focusing on participants' reactions to spoofing of the security mechanism. This paper is the first to examine trusted UI mechanisms for VR.

We identify a key contradiction---users' least preferred method was the most effective, while their most preferred method was the least effective, with a spoofed mechanism fooling 44\% of participants. This work should inform all immersive browser vendors of the risks and potential solutions that exist to secure link traversals, as well as the usability challenges of a trusted immersive UI. This paper summarizes the benefits and tradeoffs of each proposed security mechanism.

In \autoref{background}, we discuss existing 2D mechanisms for protecting against phishing attacks. Our threat model is specified in \autoref{threats} along with a definition of a trusted immersive UI, while the three privacy mechanisms we investigated are defined in \autoref{privacy-methods}. We discuss our findings (\autoref{results}) in \autoref{sec:discussion}.

\section{Background}\label{background}

Phishing attacks are a huge problem for both users and brands. A \emph{phishing} website is a fraudulent website intended to capture sensitive data such as login credentials or payment information, often accessed via a link in a targeted email. The FBI estimates that phishing attacks have cost businesses over \$12 billion from 2013 to 2018~\cite{fbi}. However, phishing sites can also be accessed via fraudulent links in websites. Often, phishing sites will look identical to the legitimate site they are imitating, with the only difference being an incorrect domain name.

TODO define origin and domain

Unfortunately, browser based phishing protection techniques have largely proven to be inadequate. Dhamija et. al. found that 23\% of users did not look at 2D indicators such as the address bar, status bar, and security indicators, leading to incorrect choices 40\% of the time~\cite{dhamija2006phishing}. They recommend placing indicators inside the users' focus of attention, as well as alerting users to untrusted states as well as trusted states.

Thompson et. al. studied the effectiveness of website identity indicators, such as Extended Validation (EV) certificates and URLs~\cite{thompson2019web}. They found that users often misidentified the origin of a URL based on the URL provided, corroborating Dhamija et. al., who showed that URLs are ineffective identity indicators. They recommended that future identity indicators incorporate user research into the design phase.

Domain highlighting is a phishing mitigation approach wherein the domain of the URL is highlighted or bolded to differentiate it from the rest of the URL. It visually enhances the domain, assuming users both recognize legitimate domain names and pay attention to the URL bar as part of their browsing. Lin et. al. examine the effectiveness of this technique in assisting users to identify phishing sites and conclude that it provides benefit to people who already attend to the URL bar; however most people do not bother to look at the address bar unless they are told to~\cite{lin2011does}.

Previously, banks deployed a system called \emph{SiteKey}. First, users entered their username and were redirected to a new page to enter their password. The password entry page also contained a unique, user-chosen image that the user was instructed to look for before entering their password as a phishing protection. Schechter et. al. found that removing this image and replacing it with an upgrade message tricked 97\% of 60 participants into entering their passwords~\cite{schechter2007emperor}.

During preliminary discussions, there were concerns that two of our proposals, the Sigil and the Agent, were similar to the \emph{SiteKey} and that users could be similarly fooled. Due to this concern, we decided to study whether participants would proceed with a link traversal if the security mechanism was replaced with a loading symbol.

Our security mechanisms are designed to indicate that the browser is rendering the content that the user is seeing---not the website. For this reason, the user can trust that the content they are seeing is accurate. For example, if a website were to render a link traversal, they could claim that the user is going to bank.com, but really send them to evilbank.com to steal their banking credentials, whereas the browser would display the ground truth---that the link is going to evilbank.com. For this reason, malicious websites may attempt to spoof our security mechanisms. Non-spoofability is the key security property that our proposals must have, and what this study examines most thoroughly.

Currently, there are no phishing mitigations designed for the immersive web, preventing immersive browsing. Instead, users must navigate using 2D mode in their immersive web browser, which is susceptible to all of the above issues. In fact, users may be more susceptible to phishing, because reading text in 2D mode can be difficult; therefore they may be fooled by small changes in domain names. Following the recommendations of Dhamija et. al., we incorporated user testing into our design phase, prior to deploying an implementation of our security mitigation. We have also placed the security indicator inside the users' attention, requiring them to interact with it before proceeding with a link traversal instead of relying on users to actively direct their attention to security indicators.

\section{Threat Model}\label{threats}
The general idea that we need to convey to users is that there are some things that the browser and the browser alone can render---the web page has no ability to draw these elements. The trust model on the web is that web pages are inherently untrustworthy. The trusted immersive UI element indicates that the user is interacting with the browser, not the webpage.

Suppose a user wishes to navigate to site B from site A by clicking on a link, but is redirected to site C. A trusted immersive UI would protect against this type of threat by displaying the true origin (site C) that a user is navigating to before navigation occurs.

Alternatively, a user might be navigated against their wish. A trusted immersive UI would protect against this threat by ensuring that the user intends on navigating to a new URL and displaying the new origin to the user prior to navigation. Essentially, the trusted immersive UI adds a small amount of friction to a link traversal in order to guarantee user intention. This is a conscious decision.

\subsection{Trusted immersive UI}
A trusted immersive UI provides mechanisms to make it clear that the user is interacting with the browser, not the content, for example when:
\begin{itemize}
    \item navigating between URLs;
    \item inspecting URLs;
    \item revoking and examining consent (accountability); and,
    \item providing user warnings such as insecure site warnings and phishing and malware protection warnings.
\end{itemize}

The following are properties of a trusted immersive UI~\cite{tiui_gh}:
\begin{itemize}
  \item non-spoofable;
  \item indicates where the request/content displayed originates from;
  \item if it relies on a shared secret with the user, the shared secret must be unobservable by a mixed reality capture;
  \item it is consistent between immersive experiences in the same UI;
  \item avoid spamming/overloading the user with prompts;
  \item easy to intentionally grant consent (e.g. the UI should be easily discovered);
  \item hard to unintentionally grant user consent (e.g. the UI should prevent clickjacking); and,
  \item provides clear methods for the user to revoke consent and verify the current state of consent
\end{itemize}

\section{Privacy Methods}\label{privacy-methods}
A key concern in our threat model is that websites will attempt to spoof our trusted UI solution; therefore, we must have a strong anti-spoofing property. For spoofing protection, there are two paths:
\begin{enumerate}
    \item A persistent UI: If spoofing occurs, it should be immediately obvious, because there will be multiple UIs shown. This requires user onboarding to train them to detect spoofing.
    \item A non-persistent UI that relies on a shared secret: The page does not know the shared secret, and therefore can not spoof the UI. This requires that the page can not capture the shared secret. It also requires that the user notice the shared secret, and notice deviations from the shared secret.
\end{enumerate}

\subsection{Logo Proposal}
\begin{figure}[ht]
    \centering
    \begin{minipage}[b]{0.45\linewidth}
    \includegraphics[width=\linewidth]{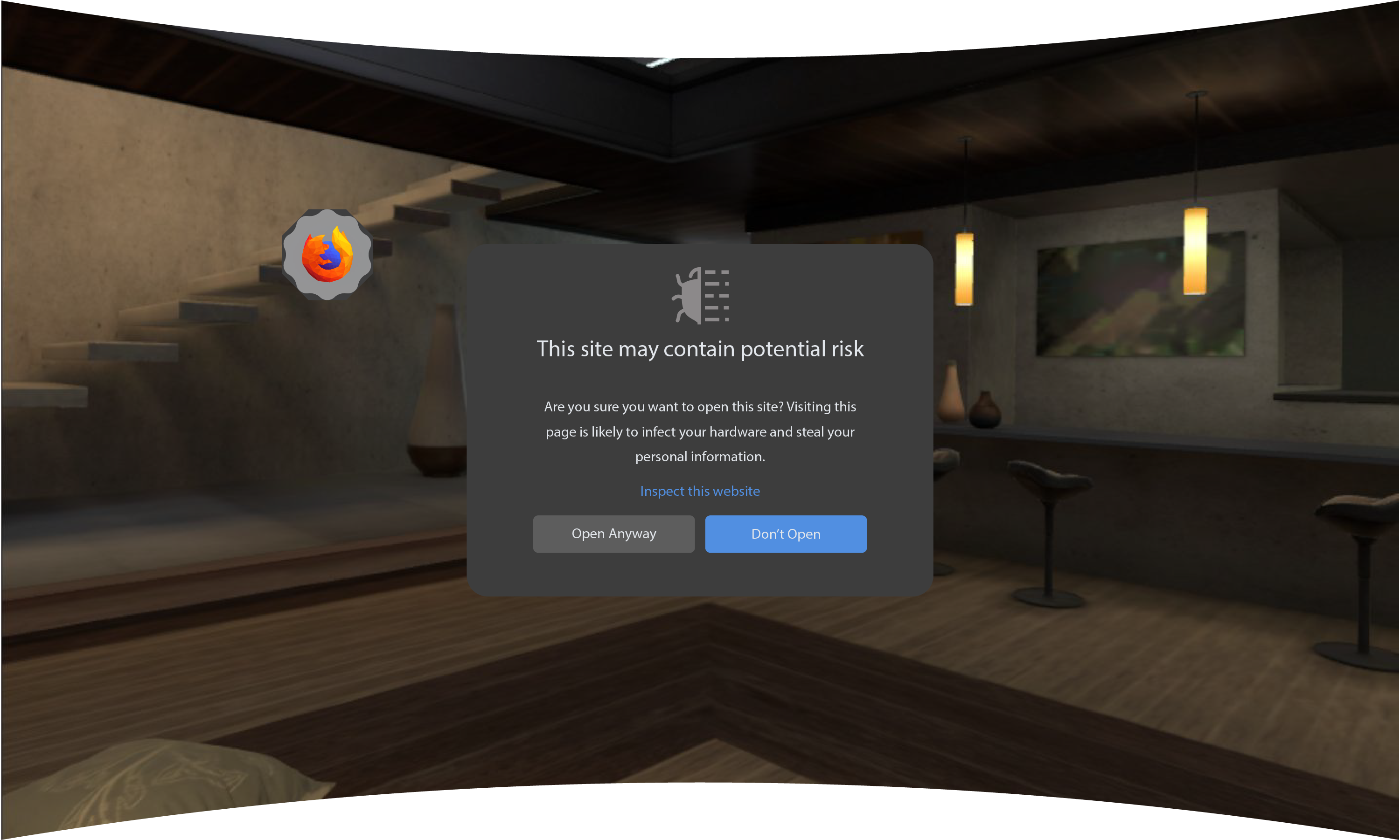}
    \caption{An activated Logo}
    \label{fig:logo}
    \end{minipage}
    \quad
    \begin{minipage}[b]{0.45\linewidth}
    \includegraphics[width=\linewidth]{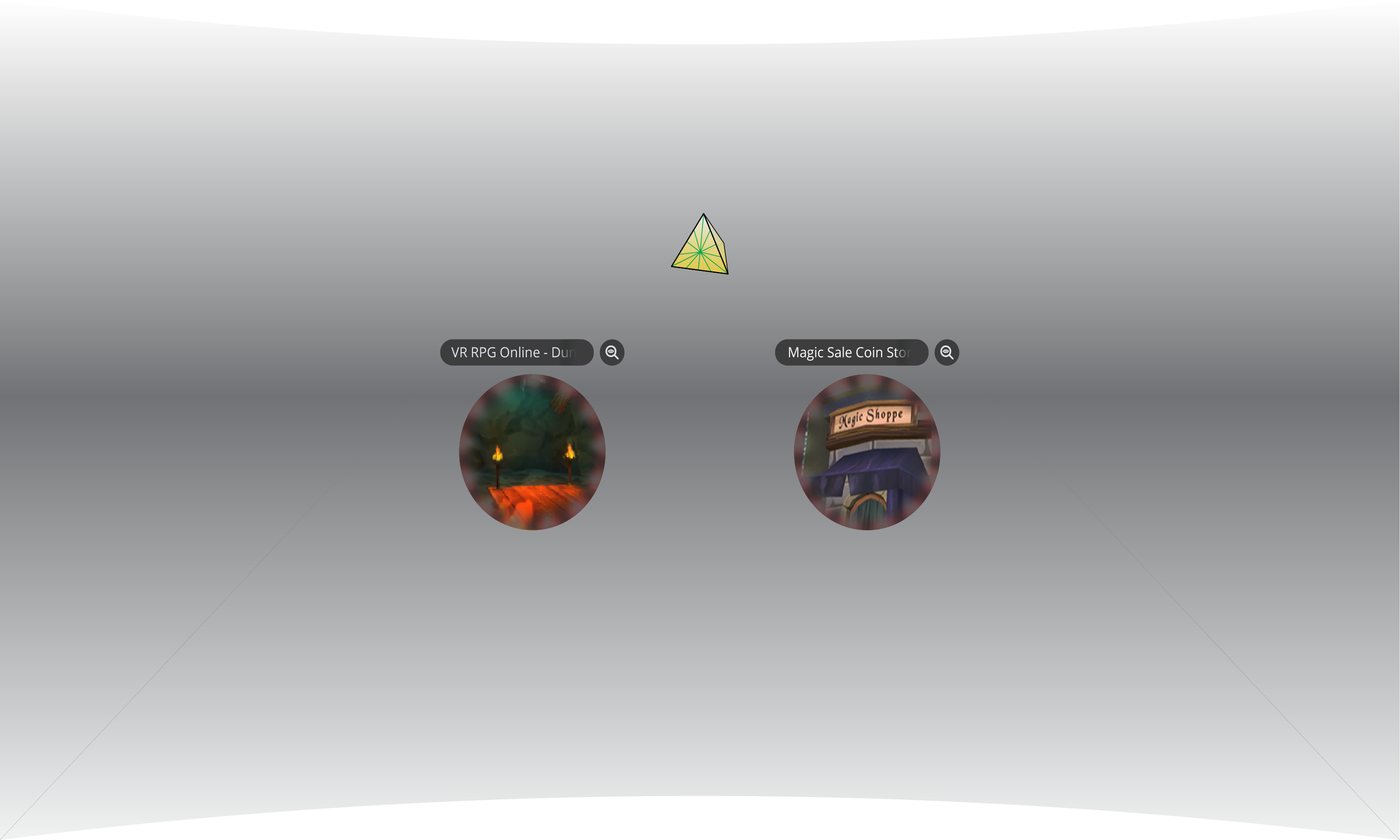}
    \caption{The authentic Sigil middle panel}
    \label{fig:sigil}
    \end{minipage}
\end{figure}

The Logo (see \autoref{fig:logo}\footnote{The Logo is an institutional logo and has been redacted for anonymization purposes}) is a persistent UI element that remains at the feet or above the head, depending on user configuration, until it is activated by either pressing the back button on the controller or clicking a link. When activated, the Logo moves to eye level, indicating that the dialog box shown is being rendered by the browser, not the website.

Because the Logo is a persistent UI element, if a website attempts to spoof it, there will be two Logos displayed---one at eye level and one either at the feet or above the head. When the Logo is activated, users should check their configured location to ensure that only one Logo is showing.

\subsection{Agent Proposal}
The Agent is a non-persistent user agent that appears when the browser, not the webpage, is rendering the content that users see. It is randomly assigned to you at startup, but it can be customized. Because of the randomness and customization, a web page cannot guess what your Agent looks like, so it forms a secret between a user and the browser.

\begin{figure}
    \centering
    \begin{minipage}[b]{0.45\linewidth}
    \includegraphics[width=\linewidth]{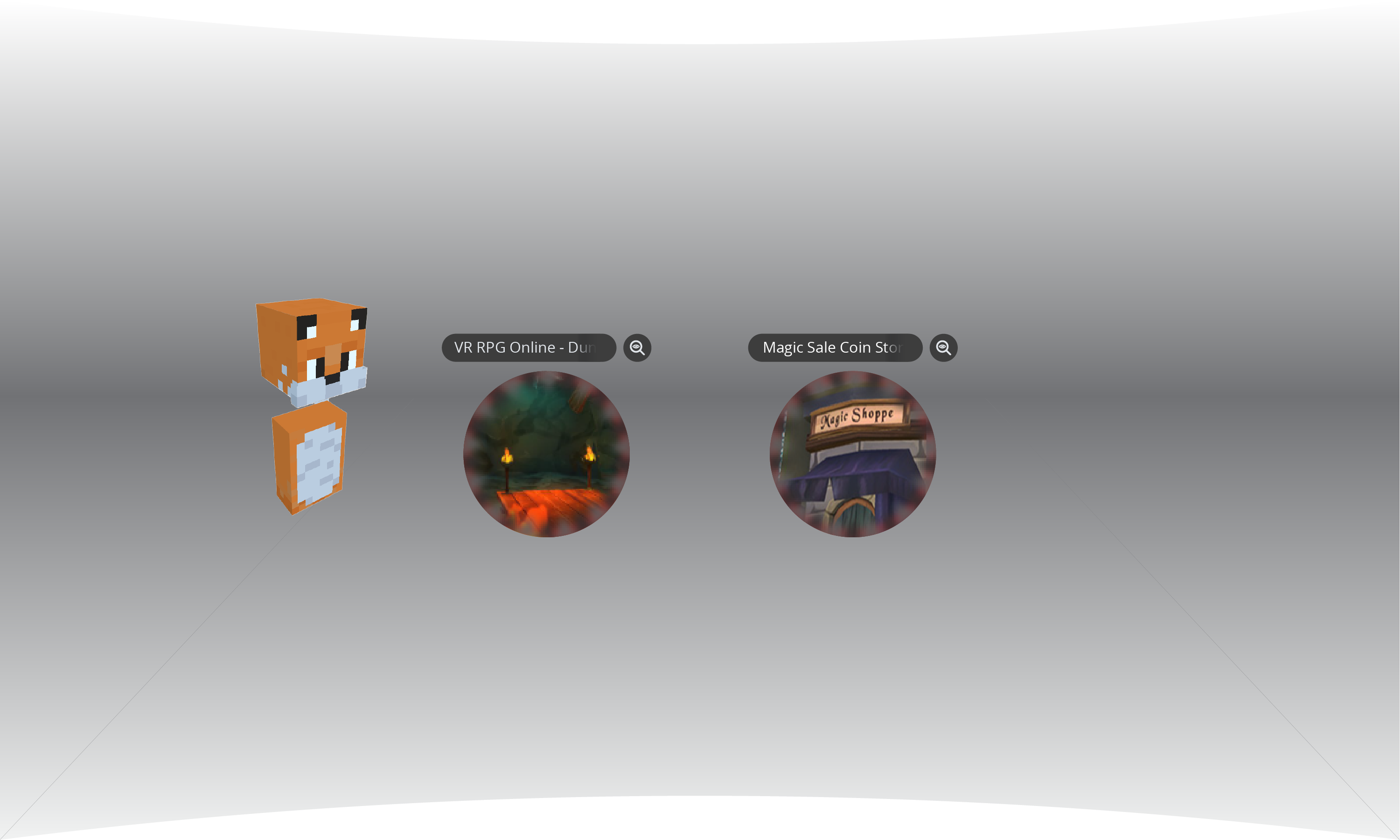}
    \caption{The authentic Agent middle panel}
    \label{fig:agent}
    \end{minipage}
    \quad
    \begin{minipage}[b]{0.45\linewidth}
    \centering
    \includegraphics[width=\linewidth]{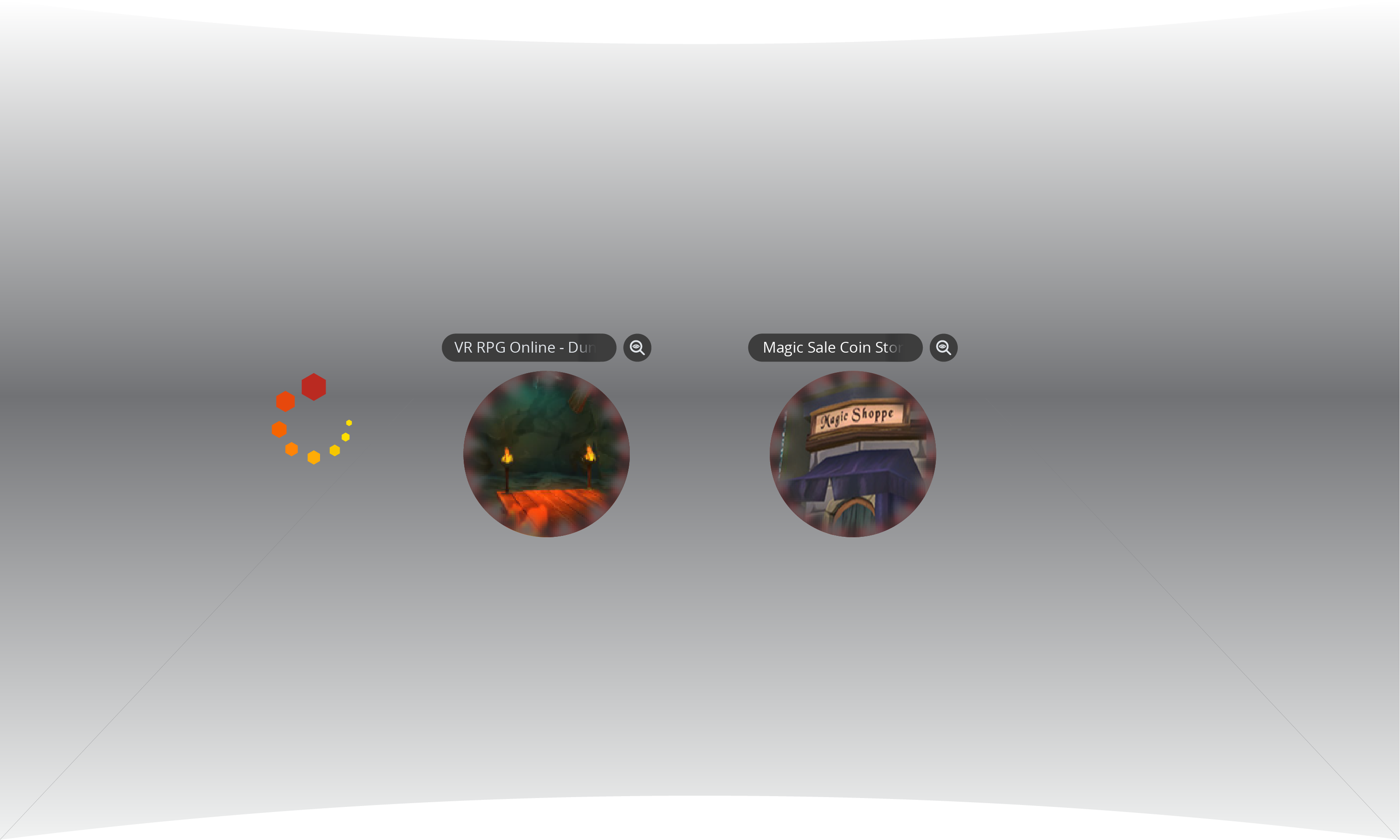}
    \caption{A loading symbol in place of an Agent}
    \label{fig:agentloading}
    \end{minipage}
\end{figure}

Upon clicking a cross-origin link, users will be redirected to an interstitial room containing their Agent and portals to their current location and the link location (\autoref{fig:agent}). Unlike other browser security mechanisms, which require that users take affirmative action, this interstitial room confronts users with the security mechanism. First, they must visually verify that the Agent that appears is the correct agent they were assigned or have customized. If it is incorrect, then the browser is not currently rendering the interstitial page and they should not continue. To verify what their Agent looks like, users can visit their browser home page.

Upon clicking a same-origin link, a dialog featuring the agent will appear asking the user if they wish to follow the link.

\subsection{Sigil Proposal}
The Sigil is very similar to the Agent, but instead of a user agent, a non-customizable symbol appears when the browser is rendering the content that appears. It is randomly assigned to users when they first start the browser. Because of this randomness, a web page cannot guess what a user’s Sigil looks like, so it forms a secret between the user and the browser. The mechanics are otherwise the same as the Agent.

\section{Methodology}\label{methodology}
Our overarching goal was to identify which of the aforementioned proposals is the most usable. We were looking for a trusted immersive UI solution with strong anti-spoofing properties (theoretically guaranteed by all three proposals) in practice. Overall, the exploration should answer two main questions:
\begin{enumerate}
    \item For each approach, do the security properties hold in practice? Is it usable? To determine usability, we asked participants to rate their likelihood of moving forward with the link traversal and asked them if they felt the concept made sense to them. Six participants provided usability scores for each concept on a scale of 1 (least usable) to 5 (most usable).
    \item Which approach do users prefer and why?
\end{enumerate}

We used Mozilla Hubs\footnote{\url{hubs.mozilla.com}} as a virtual space to host visual representations of each of the three proposals. This enabled us to immerse participants as if we had implemented each proposal. For example, instead of simply showing a picture of an avatar with the Logo at the avatar's feet, the participant takes the place of the avatar and the Logo appears to be at their feet.

At the beginning of the study, participants were introduced to the concept of navigating immersive links with a three frame story, showing the beginning, middle, and end state of traversing a link (see \autoref{fig:linktraversal}). The middle frame is where our trusted UI solution would be applied.

The participants were provided with the following scenario for deeper understanding: 
\begin{quote}
Suppose you're in a game, and you want to buy tokens; someone sends you a link---how likely are you to click the link and proceed to the store?
\end{quote}

We created four Hubs rooms: a lobby, to introduce participants to link traversal, and one room for each proposal. Participants joined a researcher in Hubs and were introduced to the concept of navigating immersive links, then introduced to each of the three proposals in a random order. First, they entered an antechamber in each Hubs room that displayed the security mechanism, and the researcher read a description of how the mechanism worked. Then the participant had the opportunity to ask any questions. Afterwards they entered the main room, where they were asked to make decisions based on their understanding of the security concept.

\begin{figure}[!htb]
\minipage{0.45\textwidth}
  \includegraphics[width=\linewidth]{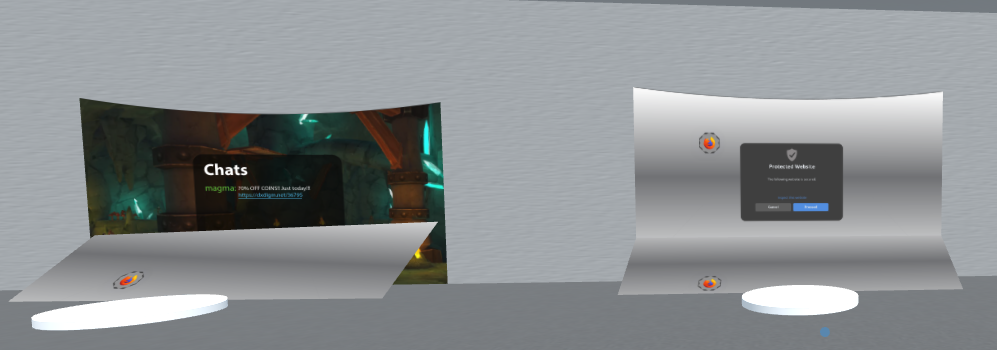}
  \caption{A frame of the Logo proposal room showing a spoofed Logo mechanism}\label{fig:logoroomspoof}
\endminipage\hfill
\minipage{0.45\textwidth}
  \includegraphics[width=\linewidth]{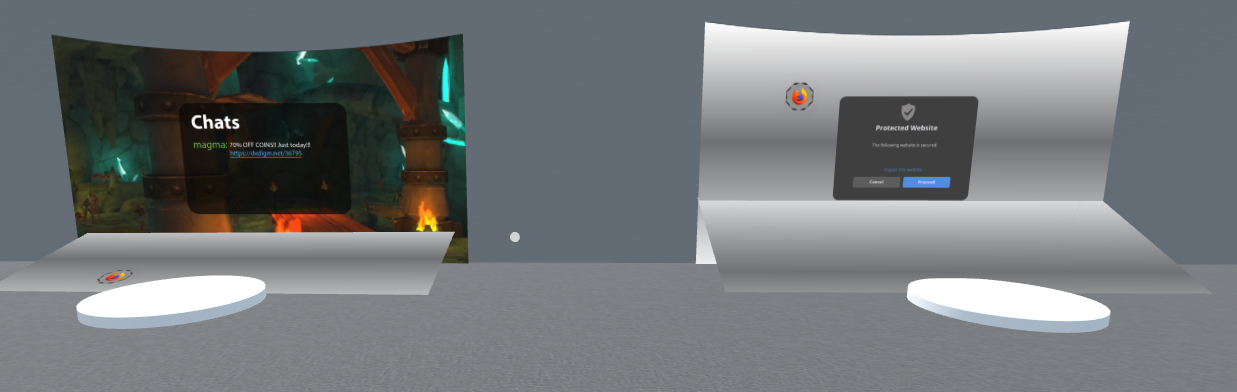}
  \caption{A frame of the Logo proposal room showing the authentic, activated Logo mechanism}\label{fig:logoroomauth}
\endminipage\hfill
\minipage{1\textwidth}
  \includegraphics[width=\linewidth]{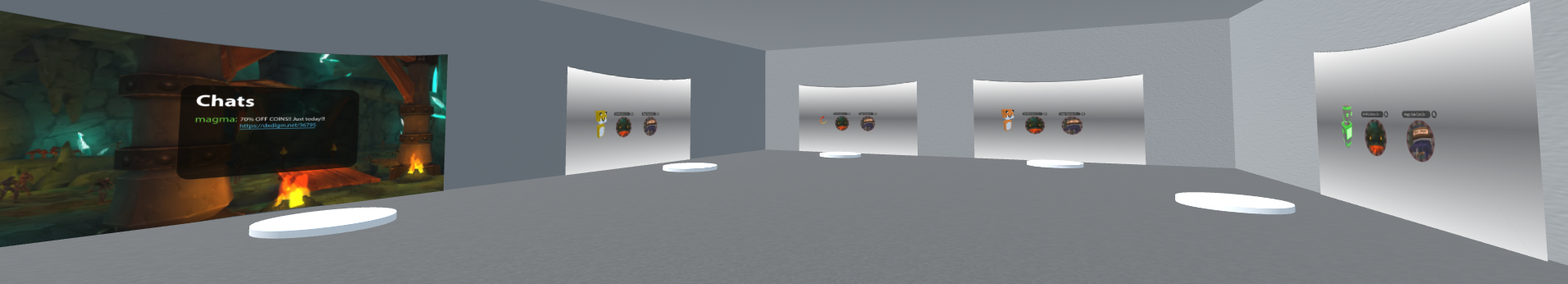}
  \caption{The Agent proposal room showing frame one of the link traversal story and the four potential middle frames shown to participants. In order, they show an Agent that is slightly wrong (the wrong color), a loading symbol, the authentic Agent, and a completely wrong Agent.}\label{fig:agentroom}
\endminipage\hfill
\minipage{1\textwidth}%
  \includegraphics[width=\linewidth]{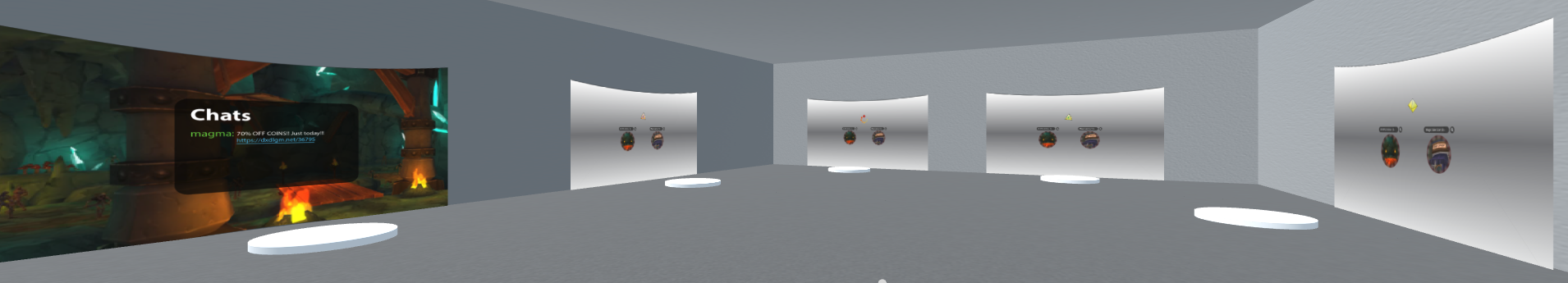}
  \caption{The Sigil proposal room showing frame one of the link traversal story and the four potential middle frames shown to participants. In order, they show a Sigil that is slightly wrong (the wrong color), a loading symbol, the authentic Sigil, and a completely wrong Sigil.}\label{fig:sigilroom}
\endminipage
\end{figure}

In the Logo room, participants viewed three link traversal stories, each with a beginning, middle, and end frame. One of these was an authentic link traversal with no spoofing (shown in \autoref{fig:logoroomauth}), while the other two displayed the two forms of spoofing possible---an additional logo either at the feet or overhead (shown in \autoref{fig:logoroomspoof}).

In the Agent room, shown in \autoref{fig:agentroom}, participants viewed one link traversal story, but with four potential middle panels. One middle panel was authentic (i.e. \autoref{fig:agent}), one was slightly wrong, one was completely wrong, and one displayed a loading symbol (see \autoref{fig:agentloading}). The order of the frames visited was not randomized.

Similar to the Agent room, the Sigil room shown in \autoref{fig:sigilroom} displayed one link traversal story with four potential middle panels. One middle panel was authentic, one was slightly wrong, one was completely wrong, and one displayed a loading symbol. The order of the frames visited was not randomized.

The researcher directed the participant to stand on the white pedestal in front of the frames to elicit a sense of immersion, just as if they were viewing a page on the immersive web. At each middle frame, the researcher asked what they saw, and on a scale of 1 (not at all likely) to 5 (extremely likely), how likely would they be to click to continue making the tokens purchase.

After seeing all the proposals, the participant was asked which they found the most usable.

\subsection{Recruitment and Demographics}
We recruited seven men and two women from the USA, Canada, Australia, and Singapore. Each participant was required to own an Oculus Quest so they could join the Hubs rooms immersively. However, two participants had technical issues and joined the Hubs rooms on their computers. We did not note any substantial differences between the participants who accessed the environment via head-mounted device or laptop.

We invited a range of participants based on the extent they reported paying attention to online security measures. Two people stated they never pay attention to online security. Four people claimed they almost always do. The remaining three participants were in the middle stating sometimes they do. We believe that recruiting people who hold different attitudes towards online security helps represent the general population of technology users.

\subsection{Ethical considerations}
Our institutions are not subject to IRB approval, but the experiment went through an internal review process. All participants signed consent forms and were again verbally consented before proceeding with the interview. Each participant received an honorarium for taking part in the study.

\section{Results}\label{results}
At the beginning of the study, we asked general questions about the relationship between the browser and websites, as well as how participants verify on the 2D web that they are on the correct webpage.

Three participants self-reported that they sometimes check the URL of a website, particularly when performing sensitive activities like banking, while one said that they know they are on the correct webpage because they type the URL themselves. Another three participants reported that they judge being at the correct location by the content of a webpage, not by any browser indicators. This corroborates prior research that indicates that users don't look at URLs and raises concerns that by properly imitating sites, phishing can be very effective~\cite{dhamija2006phishing}.

Overall, no participants mentioned that the role of the browser was to protect users' privacy and security, merely that its role is to display webpages. Two participants conflated the role of a search engine with that of a browser, claiming that the browser allows them to search for information.

We then asked participants to provide ratings from 1 (least likely) to 5 (most likely) of how likely they would be to proceed with clicking the link to proceed with the link traversal, given the security mechanism displayed. The results are shown in tables \ref{table:logo}, \ref{table:agent}, and \ref{table:sigil}. We found that having the correct shape of a Sigil or Agent, but using the wrong color was difficult for users, respectively fooling 3 and 4 participants into continuing with a spoofed mechanism.

\begin{quote}
    “This looks like my agent from earlier. But I’d give it the same amount of credibility as the yellow one. 3 out of 5. They’re pretty similar so I probably wouldn’t second guess. Maybe you could have your design next to the one popping up so you could compare them?”
\end{quote}

One possible issue for participants was that they quickly forgot what the correct Sigil or Agent looked like. One requested a 'quick-check' mechanism where she could pull up her true Sigil or Agent for a side-by-side comparison. Another stated that he couldn't be more confident until he was more familiar with his true Sigil or Agent, which he believed would come with time.

\begin{table}
\begin{tabular}{c|ccc}
     & Authentic & Overhead Spoof & Foot Spoof \\
     Participant 1 & 4 & 1 & 1  \\
     Participant 2 & 5 & 1 & 1 \\
     Participant 3 & 3 & 1 & 3 \\
     Participant 4 & 4 & 1 & 1 \\
     Participant 5 & 4 & 2 & 2 \\
     Participant 6 & 5 & 1 & 1 \\
     Participant 7 & 4 & 1 & 1 \\
     Participant 8 & 5 & 1 & 1 \\
     Participant 9 & 5 & 2 & 3 \\
     Avg Likelihood & 4.33 & 1.22 & 1.56 \\
     Median Likelihood & 4 & 1 & 1 \\
\end{tabular}
\caption{Likelihood rating of continuing with a link traversal in the Logo room}\label{table:logo}
\end{table}

\begin{table}
\begin{tabular}{c|cccc}
     & Wrong color & Loading & Authentic & Totally wrong  \\
     Participant 1 & 1 & 1 & 5 & 1 \\
     Participant 2 & 1 & 1 & 3 & 3 \\
     Participant 3 & 3 & 1 & 3 & 2 \\
     Participant 4 & 1 & 1 & 4 & 1 \\
     Participant 5 & 5 & 1 & 5 & 4 \\
     Participant 6 & 1 & 1 & 4 & 1 \\
     Participant 7 & 4 & 1 & 2 & 1 \\
     Participant 8 & 4 & 1 & 4 & 1 \\
     Participant 9 & 5 & 3 & 3 & 1 \\
     Avg Likelihood & 2.78 & 1.22 & 3.67 & 1.67 \\
     Median Likelihood & 3 & 1 & 4  & 1 \\
\end{tabular}
\caption{Likelihood rating of continuing with a link traversal in the Agent room}\label{table:agent}
\end{table}

\begin{table}
\begin{tabular}{c|cccc}
     & Wrong color & Loading & Authentic & Totally wrong  \\
     Participant 1 & 1 & 1 & 4 & 1 \\
     Participant 2 & 1 & 1 & 3 & 1 \\
     Participant 3 & 2 & 1 & 3 & 1 \\
     Participant 4 & 1 & 1 & -  & 1 \\
     Participant 5 & 5 & 2 & 5 & 1 \\
     Participant 6 & 1 & 1 & 5 & 1 \\
     Participant 7 & 1 & 1 & 4 & 1 \\
     Participant 8 & 5 & 1 & 5 & 1 \\
     Participant 9 & 5  & 3 & 4 & 2 \\
     Avg Likelihood & 2.44 & 1.33 & 4.13 & 1.11 \\
     Median Likelihood & 1 & 1 & 4 & 1 \\
\end{tabular}
\caption{Likelihood rating of continuing with a link traversal in the Sigil room.  Participant 4 did not give a numerical rating for one frame.}\label{table:sigil}
\end{table}

Tables \ref{table:logo}, \ref{table:agent}, \ref{table:sigil} show the likelihood ratings of continuing with a link traversal when presented with a frame in the Logo, Agent, or Sigil rooms, respectively. From this, we can calculate the average likelihood that a participant in the study would proceed with a link traversal when presented with an authentic, spoofed, or loading frame. As \autoref{table:logo} shows, participants had a very high likelihood of proceeding when shown only one Logo, and a very low likelihood of proceeding when shown a spoofed Logo. Likewise, \autoref{table:agent} demonstrates that participants were very unlikely to proceed when shown a loading symbol or a completely wrong Agent, but somewhat likely to proceed when shown an Agent of the wrong color. Worryingly, they were also only somewhat likely to proceed when shown their authentic Agent, showing the low confidence they had in their recall. Sigil results for the loading symbols and totally wrong Sigil were similar to that of Agent, while participants were less likely to be spoofed by a Sigil of the wrong color and more likely to proceed with their authentic Sigil (very likely versus somewhat likely).

\section{Discussion}\label{sec:discussion}

Overall 6 out of 9 (67\%) participants preferred the Agent concept, while the other 3 preferred the Sigil. Some perceived positives of the Agent versus the Sigil were the customization possibilities, with participants believing that customization would help them remember their Agent. However, they were also concerned that people would pick similar customizations, presenting a security risk, or potentially forget the customizations they made, especially if they frequently changed their customizations. One participant proposed adding a customizable name to the Agent that would appear as an additional authentication factor.

Only two professed any preference for the Logo, calling it easy to use. This suggests that the description of the Logo concept was perhaps inadequate and more work on onboarding users is required.

\begin{quote}
    “It’s more complex to understand. While we’re here, I can ask questions and take too much time. But most users who aren’t familiar with internet security won’t understand the concept and won’t know when to trust situations. 2 or 3. Is there an easy way to teach the audience about this?”
\end{quote}

However, another user believed that this concept would easily be grasped by a majority of users, despite preferring another method:

\begin{quote}
    I hand my headset to kids all the time...it feels like <the Logo> is gonna be the easiest to explain to someone like them... I'll just use my parents as the example. It's really easy to say `Look for two---anytime you see one, look for two. If you've got one, then it's OK!' And that's really straightforward." % - Participant 4
\end{quote}

\subsection{Loading symbol}
Due to the findings of Schechter et. al~\cite{schechter2007emperor}, we were particularly interested in what users would think when confronted with a loading symbol instead of the security mechanism. Numerous participants needed to be prompted to consider the symbol as a loading symbol, because it was not animated, which is a potential limitation of this study.

Only one participant confessed that they were impatient and likely to click through before the loading symbol resolved into the security mechanism (either Sigil or Agent). Most participants said that they would wait for the loading to resolve or reload the page. 

\begin{quote}
    “Either it’s trying to load my agent or it isn’t secure and this is a red flag. If it was the loading screen I was used to seeing, then 3 out of 5. Maybe I would trust it, but maybe I would wait.” 
\end{quote}

One participant correctly identified that there was no reason for there to be a loading symbol. The security mechanisms are browser-based and should be stored on the device, so there should be nothing to load over the network.

\subsection{Limitations and Challenges}
This study evaluated participants' recall of three security mechanisms over a short period of time. In a real world scenario, it is possible that users could have long breaks between using their immersive devices and need to recall a security mechanism over long periods of time. While the Sigil and Agent mechanisms provide a way to verify what your correct Sigil or Agent look like, deployments may want to provide a 'refresher course' in the functioning of the mechanism as a whole.

Whichever trusted immersive UI method a platform chooses to deploy will require an onboarding and educational component. For example, the customization process for the Agent concept could be used to introduce users to the problem and mitigation, instructing them to look for their agent during security sensitive tasks such as link traversals. For the Logo and Sigil concepts, a separate onboarding experience would need to be created and deployed when the feature is launched. An additional complication is that all users in this study share their devices, presenting a challenge for users who have been handed a headset but not appropriately inducted into the security mechanism meaning and working. Further work should explore how casual users can safely traverse links in a shared user model.

While we randomized the order of the concepts users viewed, we did not randomize the order of the frames presented in the rooms, which could influence participants' reactions and ratings.

\subsection{Future Research Directions}
During the course of the study, we identified six promising future research directions.

\textbf{Inspecting URLs}: In the Logo interstitial screen, participants were presented with an option to 'inspect' a URL prior to visiting it, which intrigued them. More research is needed to determine what information would be actionable for users in deciding whether to move forward with a link traversal, or if presenting this option increases the cognitive burden on users and should not be included.

\begin{quote}
    “I see the logo… not sure if it’s over my head or at eye level. I wish I could see ‘inspect this website.’ Do I know that this is my dialogue that comes up from my browser? If it's what I’m used to seeing, I’d inspect it probably. And curious what it would show me. I would like to see the website’s credentials. If I inspect the website and it looks good, I’d be likely to proceed. 4 out of 5”
\end{quote}

\textbf{Education methods}: Any platform choosing to deploy a trusted immersive UI should investigate the most effective way to educate users on their chosen method. A particular challenge will be educating secondary users in the functioning of the security mechanism.

\textbf{Customization}: The Agent concept allows customization. Any platform choosing to implement the Agent mechanism should investigate the effects of customization on user recall and spoofing, as well as the ability for the customization process to serve as an education method.

\textbf{Dimensions of randomness and change}: What dimensions of change (e.g. color/shape) are more difficult to detect to design effective Sigils or Agents? Additionally, how many dimensions of randomness are required for security properties to hold?

\textbf{AR contexts}: Finally, this paper is focused on exploring trusted immersive UI for virtual reality devices, but the immersive web is also designed to work for augmented reality devices. Further research should explore how to adapt the concepts presented here to AR contexts.

\textbf{Accessibility}: The concepts presented in this paper require users to visually verify security mechanisms. Further work is required to identify mechanisms that will be accessible for the visually impaired.

\section{Conclusion}\label{conclusion}
In this paper, we present the results from a 9-person study to investigate the efficacy and usability of three security mechanisms for securing link traversals on the immersive web against unintended navigations or redirection attacks. In particular, we investigated the effect of a loading screen on two security mechanisms, and found there to be little negative impact. A contradiction that we identify is that users' least preferred method was the most effective, while their most preferred method was the least effective. Based on these results, we recommend that systems implementing open-ended navigation in immersive contexts should use the Logo approach which functions well with the current shared device paradigm. Additional research into efficiently onboarding and educating users will be essential to the success of the deployment of the Logo method. While this method was the least preferred, we believe that with improved onboarding and education, this method will be easily understood and grasped by a majority of users. We also present six promising research directions identified from the study.

\bibliographystyle{ACM-Reference-Format}
\bibliography{sample-base}

\end{document}